\documentclass[twocolumn]{aa} 
\usepackage{graphicx}
\usepackage{txfonts}
\usepackage{lscape}
\usepackage{epsfig}
\usepackage{times}
\usepackage{graphicx}
\usepackage{natbib}
\bibpunct{(}{)}{,}{a}{}{}
\begin{document}
   \title{Comparing dynamical and photometric-stellar masses
   of early-type galaxies at $z\sim 1$ \thanks{Observations have been
   carried out using the Very Large Telescope at the ESO Paranal
   Observatory under Program IDs: LP168.A-0485, 169.A-0458,
   170.A-0788, 70.A-0548} }

   \author{A. Rettura\inst{1,2,3} \thanks{Present address: Department of Physics and Astronomy, Johns Hopkins University, 3400 N. Charles Str., Baltimore, MD 21218, USA }
          \and 
	  P. Rosati\inst{1}
	  \and
          V. Strazzullo \inst{1,6}
          \and
	  M. Dickinson \inst{4}
          \and
          R.A.E. Fosbury \inst{5}
          \and
          B. Rocca-Volmerange \inst{2,3} 
	  \and
	  A. Cimatti \inst{7}
	  \and
	  S. di Serego Alighieri \inst{7}
	  \and
	  H. Kuntschner \inst{5}
	  \and
	  B. Lanzoni \inst{8}
	  \and
	  M. Nonino \inst{9} 
	  \and
	  P. Popesso \inst{1,5}
	  \and
          D. Stern \inst {10}
	  \and
	  P.R. Eisenhardt \inst{10}
	  \and
	  C. Lidman \inst{11}
	  \and
	  S.A. Stanford.\inst{12,13} 
          }

   \offprints{A. Rettura, arettura@eso.org}

   \institute{European Southern Observatory, Karl Schwarzschild
         Strasse 2, Garching bei Muenchen, D-85748, Germany \and
         Universit\'e Paris-Sud 11, Rue Georges Clemenceau 15, Orsay,
         F-91405, France \and Institut d'Astrophysique de Paris,
         UMR7095 CNRS, Universit\'e Pierre \& Marie Curie, 98bis Bd
         Arago, 75014 Paris, France \and NOAO, 950 North Cherry
         Avenue, P.O. Box 26732, Tucson, AZ 85726-6732, United States
         \and ST-ECF, Karl-Schwarzschild-Strasse 2, D-85748 Garching,
         Germany \and Dipartimento di Scienze Fisiche, Universit\`a
         degli Studi di Napoli ``Federico II'', via Cinthia, I--80126
         Napoli, Italy \and Istituto Nazionale di Astrofisica (INAF),
         Osservatorio Astrofisico di Arcetri, Largo E. Fermi 5, 50125
         Firenze, Italy \and Istituto Nazionale di Astrofisica (INAF),
         Osservatorio Astronomico di Bologna, via Ranzani 1, 40127
         Bologna, Italy \and Istituto Nazionale di Astrofisica (INAF),
         Osservatorio Astronomico di Trieste, via G. B. Tiepolo 11,
         I-34131 Trieste, Italy \and Jet Propulsion Laboratory,
         California Institute of Technology, 4800 Oak Grove Drive,
         Pasadena, CA 91109; United States \and European Southern
         Observatory, Alonso de Cordova 3107, Vitacura, Casilla 19001,
         Santiago 19, Chile \and Department of Physics, University of
         California at Davis, 1 Shields Avenue, Davis, CA 95616-8677;
         United States \and Institute of Geophysics and Planetary
         Physics, Lawrence Livermore National Laboratory, L-413,
         P.O. Box 808, 7000 East Avenue, Livermore, CA 94551; United
         States }


  \abstract 
  {} 
  {The purpose of this study is to explore the relationship
  between galaxy stellar masses, based on multiwavelength
  photometry spectral template fitting and dynamical masses based on
  published velocity dispersion measurements, for a sample of 48 early-type
  galaxies at $z \sim 1$ with \textit{HST}/ACS morphological information.}  
  {We determine photometric-stellar masses and perform a quantitative
  morphological analysis of cluster and field galaxies at redshift
  $0.6< z < 1.2$, using ground- and space-based multiwavelegth data
  available on the GOODS-S field and on the field around the X-ray
  luminous cluster RDCS1252.9-2927 at $z=1.24$. We use multi-band
  photometry over 0.4-8$\mu$m from \textit{HST}/ACS,
  \textit{VLT}/ISAAC and \textit{Spitzer}/IRAC to estimate
  photometric-stellar masses using Composite Stellar Population (CSP)
  templates computed with PEGASE.2 (Fioc \& Rocca-Volmerange, 1997)
  models. We compare stellar masses with those obtained using CSPs
  built with Bruzual \& Charlot (2003; BC03) and Maraston (2005; M05)
  models.  We then compare photometric-stellar mass and dynamical mass
  estimates as a function of morphological parameters obtained from
  \textit{HST}/ACS imaging.} 
  {Based on our sample, which spans the mass range $\log M_{phot}
  \simeq [10,11.5]$, we find that 1) PEGASE.2, BC03, M05 yield
  consistent photometric-stellar masses for early-type galaxies at $z
  \sim 1$ with a small scatter (0.15 dex rms); 2) adopting a Kroupa
  IMF, photometric-stellar masses match dynamical mass estimates for
  early-type galaxies with an average offset of 0.27 dex; 3) assuming
  a costant IMF, increasing dark matter fraction with the increasing
  galaxy mass can explain the observed trend; 4) we observe that
  early-type galaxies with significant disk components (Sa/Sab) or
  with signs of dynamical interaction tend to have the largest
  deviation from a one-to-one $M_{dyn} vs M_{phot}$ relation.}
  {}

\keywords{galaxies: clusters: individual: RDCS J1252-2927 - galaxies: evolution - galaxies:
formation - galaxies: elliptical and lenticular, cD - galaxies:
kinematics and dynamics - cosmology: observations.}

   \titlerunning{Dynamical and photometric-stellar galaxy mass estimates at $z\sim 1$}
   \maketitle
%

\section{Introduction}
Galaxy surveys have traditionally built samples based on the
luminosities of galaxies, and have used luminosity functions to study
the evolution of the galaxy population. The need to select galaxies on
the basis of their (stellar or possibly total) mass and to trace the
evolution of the space density of galaxies in mass bins has long been
advocated. In such a way, one can reconstruct the build-up of galaxy
mass across cosmic time and directly compare observations to
predictions of currently favored hierarchical structure formation
models. The evolution of the galaxy luminosity function is generally
affected in a complex fashion by all the physical processes which
modulate the star formation history of galaxies, whereas the mass
function evolves in a smoother way and is a direct probe of galaxy
formation models (e.g., \citet{Baugh03}, \citet{Hernquist03},
\citet{Somerville04}, \citet{Nagamine04}, \citet{DeLucia05}). This has
motivated many authors during the last decade to carry out surveys by
selecting galaxies at near-infrared rest-frame wavelengths (e.g.,
\citet{Songaila94}, \citet{Cohen99}, \citet{Drory01},
\citet{Cimatti02}), since the rest-frame near-IR light is closely
related to the total mass (e.g., \citet{Gavazzi96}). \\More
recently, the efficiency with which multi-wavelength photometry can be
collected from ground-based and space-based observatories has enabled
the measurement of stellar masses by fitting widely-sampled spectral
energy distributions (SED) of galaxies with spectrum synthesis models
(e.g., \citet{Brinchmann00}, \citet{Cole01}, \citet{Papovich01},
\citet{Shapley01}, \citet{Dickinson03}, \citet{Fontana03},
\citet{Rudnick03}, \citet{Fontana04}, \citet{Drory04a},
\citet{Saracco04}, \citet{Rocca04}).  In parallel, considerable
efforts have been devoted to measuring the evolution of the
mass-to-light ratio of galaxies out to $z\sim 1$, via Fundamental
Plane studies (FP, \citet{Djorgo87}, \citet{Dressler87}) at different
environmental densities (e.g.  \citet{Franx93}, \citet{vandokkum96},
\citet{Busarello97}, \citet{vandokkum98}, \citet{Treu01},
\citet{Gebhardt03}, \citet{Treu05}, \citet{vdw05a},
\citet{dSA05}). These studies have provided increasing evidence that
galaxy mass plays a critical role in driving galaxy evolution.  While
it has become common to derive stellar masses from multi-wavelength
surveys via SED fitting methods (the so called photometric-stellar
masses), it remains important to test the reliability of this
methodology which relies heavily on stellar population spectrum
synthesis models.\\
Moreover, by comparing dynamical and photometric-stellar
masses one can investigate the relevance of the dark matter component
of early-type galaxies as a function of the total mass (e.g.,
\citet{Napolitano05}).\\
\citet{Drory04b} have performed such a comparison for a large sample
of local galaxies in the SDSS \citep{York00} spanning a large mass
range, $8 < log(M_{phot}/M_{\odot}) < 12$.  In this paper, we extend
their study to higher redshifts, inevitably focusing on massive
systems with $M_{phot} \ge 10^{10} M_{\odot}$. For these galaxies,
\citet{Drory04b} find a good match between the photometric-stellar
mass, $M_{phot}$, and a quantity $M_{dyn }\propto
\frac{\sigma_{0}^{2}R_e}{G}$, although with a substantial scatter
around the line of equality.\\ In a similar study, \citet{Lintott05}
have compared dynamical and photometric-stellar masses within the
effective radii of a sample of local massive ($M_{phot} > 8 \cdot
10^{10} M_{\odot}$) early-type galaxies also selected from the SDSS.
They find $M_{dyn} \propto M_{phot}^{1.3}$, suggesting that dark
matter might become increasingly important in massive galaxies. \\
\citet{Cappellari05} have studied the correlation between the
dynamical $M/L$ from virial and Schwarzschild modeling and the
photometric-stellar mass (all of them within the effective radius) for
a sample of early-type galaxies at $z \sim 0$, with high $S/N$
integral field spectroscopy.  They use \citet{VZ96,VZ99} stellar
synthesis models and assume a \citet{Kroupa01} IMF to compute
photometric-stellar masses. They find the adoption of a Salpeter IMF
to provide unphysical results, as a number of galaxies would have
$M_{phot} > M_{dyn}$ and note that $\sim 30 \%$ dark matter within the
effective radius and/or IMF variations are needed to explain the
discrepancies between the two mass estimators.  This is consistent
with earlier findings at $z \sim 0$ from dynamical studies (e.g.,
\citet{Gerhard01}, \citet{Thomas05}) and at $z \sim 1$ from
gravitational lensing (e.g., \citet{Treu04}, \citet{Rusin05}).\\ Using
total masses derived from gravitational lensing study of a sample of
$0.3 < z < 1.0$ early-type galaxies, \citet{Ferreras05} note a
transition from little or no dark matter in the inner regions (within
the effective radius) to dark matter dominating in the outer regions
in the massive early-type galaxies, whereas no such a trend is
observed in lower mass galaxies. \\
On a sample of 17 early-type galaxies at $z \sim 1$ \citet{dSA05} find a
good agreement between the two mass estimators for the high mass
galaxies, but most of their lower mass galaxies have stellar masses
larger than the dynamical ones.\\ Here, we significantly extend their
work {\it i}) by using a sample almost a factor of 3 bigger, {\it ii})
by measuring in a homogeneous fashion structural parameters for the
entire galaxy sample using deep \textit{HST}/ACS data, and {\it iii})
by including rest-frame near-infrared photometry to estimate
photometric-stellar masses.  With the availability of
\textit{Spitzer}/IRAC (InfraRed Array Camera) data \citep{Fazio04},
one can access the critical restframe near IR-domain
at $z\simeq 1-2$, thus enabling a more accurate determination of
photometric-stellar masses at redshifts where a significant fraction
of the stellar mass is being assembled (e.g., \citet{Dickinson03}).\\
This paper is organized as follows. In Sect. 2 we describe the
sample selection; in Sect. 3 we present the dataset and describe the
data analysis; in Sect. 4 we describe the morphological analysis based
on \textit{HST}/ACS images; in Sect. 5 we account for our method of
deriving photometric-stellar masses for different spectrum synthesis
models and discuss the results of their comparison; in Sect.6 we
derive dynamical masses and discuss their comparison with
photometric-stellar estimates in sect. 7. And finally in Sect. 8 we
summarize the results.\\

We assume a $\Omega_{\Lambda} = 0.73$, $\Omega_{m} = 0.27$ and $H_{0}
= 71\ \rm{km} \cdot \rm{s}^{-1} \cdot \rm{Mpc}^{-1}$ flat universe
\citep{Spergel03}, and use magnitudes in the AB system throughout this
work.
   \begin{figure}[ht!]
   \centering
   \includegraphics[width=8cm]{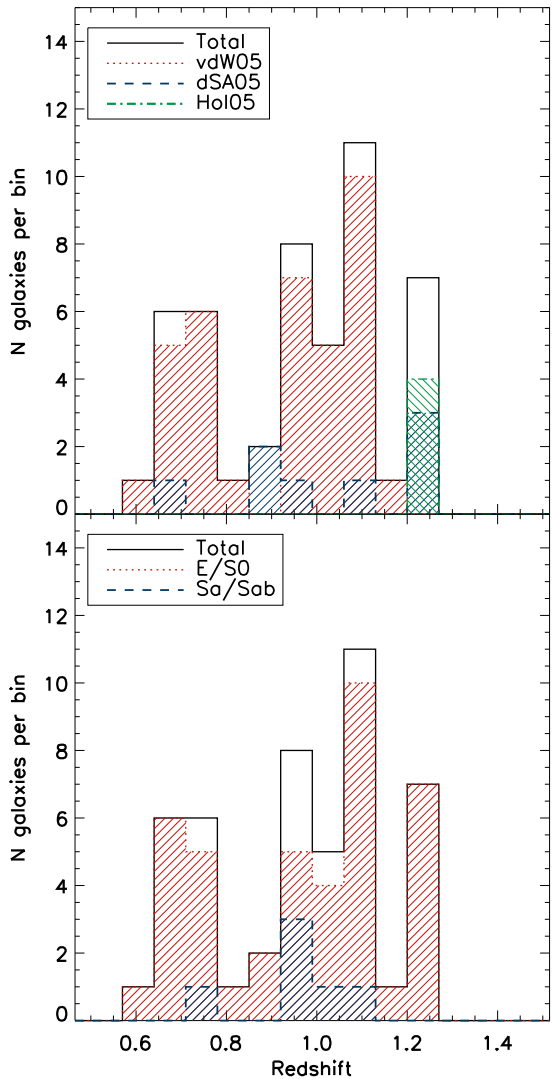}
       \caption{Top panel: The galaxy redshift distribution of the
              whole sample, consisting of 48 galaxies (black line), is
              compared with the distribution of each sub-sample
              over-plotted in red, blue and green for vdW05, dSA05 and
              Hol05, respectively.  Bottom panel: galaxy redshift
              distribution divided by morphological classes, 42
              early-types (E/S0) (red) and 6 bulge-dominated spirals
              (Sa/Sab)(blue). Images of these 6 early-type spiral
              galaxies are shown in Fig. ~\ref{spirals}.}
         \label{zspec}
   \end{figure}

\section{Sample selection}

This work is based on imaging data from two fields which have
extensive multiwavelength coverage over $0.4-8\mu$m from a combination
of high-quality deep imaging with \textit{HST}/ACS (Advanced Camera
for Surveys), \textit{VLT}/ISAAC, and \textit{Spitzer}/IRAC: the Great
Observatories Origin Deep Survey, GOODS-South field \citep{Giava04}
and the field around the cluster RDCS1252.9-2927 at $z=1.24$
\citep[hereafter Cl1252;][]{Rosati04}.  In addition, accurate velocity
dispersions of mostly early-type galaxies at $z\sim\! 1$ have been
published from studies of the Fundamental Plane in these two fields
\citep{vdw05a,Hol05,dSA05}, using deep \textit{VLT}/FORS2
spectroscopy.  Therefore, this data set provides a sample of distant
galaxies spanning an adequate mass range, with high-quality
morphological information, which is very well suited for a comparison
of stellar-photometric masses with dynamical
estimates.\\
Specifically, we selected
\begin{itemize}
\item{27\footnote{The vdW05 and dSA05 samples have two sources
in common in the CDFS with slightly different published measurements
of $\sigma_0$.  In this work, we adopt the values published
in dSA05 as the spectroscopic data were also in hand.} and 9 galaxies
in the CDFS and CL1252 fields respectively from the work of \citet[][
hereafter vdW05]{vdw05a},}
\item{8 CDFS galaxies from \citet[][ hereafter dSA05]{dSA05},}
\item{4 cluster member galaxies of Cl1252 from \citet[][ hereafter
Hol05]{Hol05}}.
\end{itemize}
These targets were either color- (vdW05, Hol05) or spectroscopically-
(dSA05) selected as early-type galaxies spanning a redshift range $0.62<z<1.24$
(see Figure~\ref{zspec}). As is customary in FP studies, velocity
dispersions were measured in a normalized circular aperture with a
diameter of  $2 R_J= 1.19 \mbox{h}^{-1} \mbox{kpc}$, equivalent to $3.4''$ at
the distance of the Coma cluster, as described by
\citet{Jorgensen95}.\\

Hence, our final sample consists of 48 galaxies with
$<z>\simeq 1$, with accurate velocity dispersions derived from
\textit{VLT}/FORS2 spectroscopy. The reader is referred to the
aforementioned papers for object selection and coordinates and for a
detailed description of the $\sigma_0$ measurements. Galaxies in this
sample have $H$ absolute rest-frame magnitudes in the range $-21.6 <
M_{H} <-24.67$, and rest-frame $U-V$ colors $1.35<U-V<1.99$.\\

\section{Cataloging and Data Analysis}
In order to accurately study both the morphology and the SED of our
galaxy sample at $z\sim 1$, a deep and homogeneous multiwavelength
dataset is needed.\\ The availability for both fields of deep
\textit{HST}/ACS imaging gives access to high quality galaxy
morphologies and makes this dataset ideal for performing a reliable
quantitative analysis. In order to obtain the morphological parameters
in the restframe B-band (often used as a reference framework for
morphological studies), we have used deep \textit{HST}/ACS images
taken with the $F850LP$ filter for both fields. We expect the
morphological k-correction, from restframe V- (for the lower redshift
galaxies of our sample) to B-band, on the derived scale parameters to
be small, in particular for early-type galaxies (e.g.,
\citet{Bohlin91}, \citet{Giava96}, \citet{Kuchinski00},
\citet{Papovich03}). \\ To build SEDs of galaxies of the GOODS-S
field, we have used optical \textit{HST}/ACS ($B_{F435W}$,
$V_{F606W}$, $i_{775W}$, $z_{F850lp}$) \citep{Giava04},
\textit{VLT}/ISAAC near infrared ($J$,$K_s$) (Vandame et al, in prep.)
and \textit{Spitzer}/IRAC ($3.6 \mu\mbox{}m$, $4.5 \mu\mbox{}m$, $5.8
\mu\mbox{} m$, $8.0 \mu\mbox{}m$) photometry (Dickinson et al., in
prep.) which is publicly available through the GOODS collaboration
\footnote{See http://www.stsci.edu/science/goods/}. \\ For field and
cluster galaxies in the Cl1252 field, we have used ground-based
optical \textit{VLT}/FORS2 ($B$, $V$, $R$), space-based optical
\textit{HST}/ACS ($i_{F775W}$,$z_{F850lp}$) \citep{Blake03}, near
infrared \textit{VLT}/ISAAC ($J_s$,$K_s$) \citep{Lidman04} and
\textit{Spitzer}/IRAC ($3.6 \mu\mbox{}m$, $4.5 \mu\mbox{}m$)
photometry (Stanford et al.,in prep.).

Accurate PSF-matched photometry, i.e. photometric measurements
normalized to the same aperture and angular resolution, is essential
to build unbiased galaxy SEDs.  To account for the large variations of
the PSF throughout our dataset (from $\simeq 0.1\arcsec$ FWHM of
\textit{HST}/ACS to $\sim\!2\arcsec$ of \textit{Spitzer}/IRAC), we
first perform photometry in $3 \arcsec$ diameter apertures in each
pass-band, we then apply aperture corrections out to $7 \arcsec$
radius. The latter are derived from a growth curve analysis of
point-like sources (identified in the ACS images) in each
passband. This approach is preferable to the one of smoothing all
images to the worst angular resolution which can result in significant
source blending. We note that the adopted aperture corrections
out to a radius of $7 \arcsec$ allow more than $95\%$ of the galaxy light to be
recovered.

   \begin{figure}[t!]
   \centering
   \includegraphics[width=8cm]{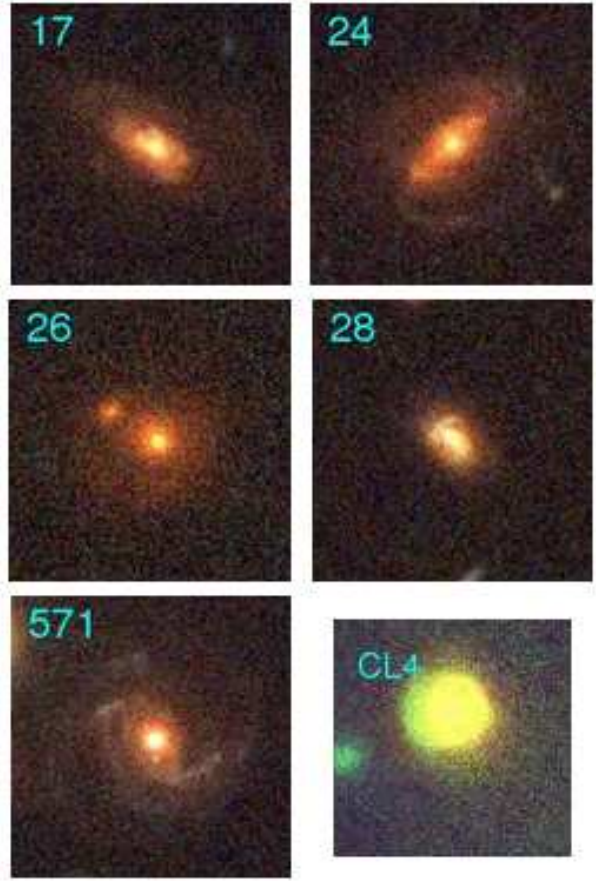}
      \caption{Color ($B, V, i, z$) images ($5\arcsec$ across) of
              visually classified (Sa/Sab) early-type-spiral
              galaxies. The morphological analysis reveals the
              presence of an underlying stellar disk and/or a clear
              spiral-arm pattern for these sources. Shortened labels
              indicate objects CDFS-17, CDFS-24, CDFS-26, CDFS-28,
              CDFS-571 and CL1252-4 in our catalogue. Note that the
              set of color we use for CL4 is a different one ($B, z, Ks$).
              We point out that the green crescent in CL4 is possibly a
              morphologically disturbed spiral arm.}
         \label{spirals}
   \end{figure}

   \begin{figure}[t!]
   \centering
   \includegraphics[width=8cm]{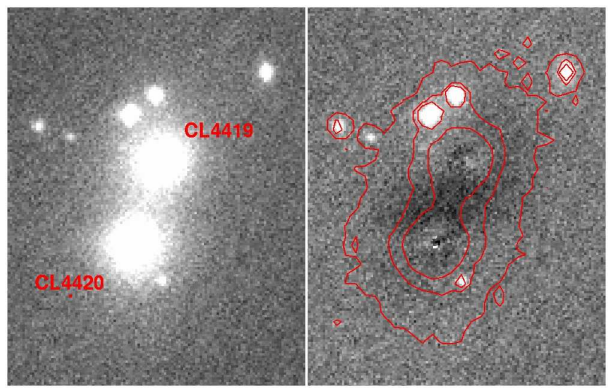}
      \caption{Left: ACS $z_{850lp}$ image of the central bright cluster
      galaxies (BCGs) of the X-ray luminous cluster CL1252 at
      $z=1.24$. Right: Best-fit S\'ersic-model-subtracted image of the
      same region of CL1252 which reveals signs of interaction in the
      form of an S-shaped residual linking the two galaxy centers.}
         \label{core}
   \end{figure}

 \begin{figure}[h]
   \centering
   \includegraphics[width=8cm]{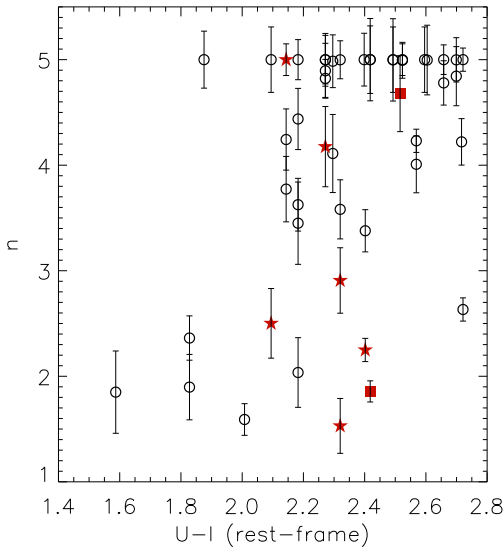}
      \caption{The correlation of the S\'ersic index, $n$, with the
      $U-I$ rest-frame color (AB magnitudes). Filled red stars symbols
      indicate the bulgy-spirals subsample (see Fig. ~\ref{spirals}).
      Filled red squares indicate $CL4419$ and $CL4420$, the pair of
      BCGs of the cluster CL1252 at $z=1.24$ showing evidence of
      mutual dynamical interaction (see also Fig. ~\ref{core}). }
 \label{ncolor}
   \end{figure}


\section{\textit{HST}/ACS morphologies of $z \sim 1$ early-type galaxies}

A description of the surface brightness (SB) distribution
of early-type galaxies is provided by the `S\'ersic $R^{1/n}$'
profile \citet{Sersic68}, with more luminous galaxies having larger
S\'ersic indices $n$, i.e., steeper light distribution towards the
center \citep{Caon93, Donofrio94, Graham96}. We have used
GIM2D, a fitting algorithm for parameterized two-dimensional modelling
of SB distribution \citep{Simard98,Marleau98} to fit each galaxy light
distribution of our entire sample with a \citet{Sersic68} profile of the
form:
\begin{equation}  \label{eq:sersic}
I(r)=I_{e_{n}}\cdot10^{-b_n[(r/R_{e,n})^{1/n}-1]},
\end{equation}
where $b_n = 1.99 n -0.33$ \citep{Cap89}, and $R_{e,n}$ is the
effective radius (i.e., the projected radius enclosing half of the
light). The classical de Vaucouleurs profile thus simply corresponds
to $n=4$ and $b_n=7.67$ in eq.(1). \\ In this work, we allow $n$ to
span the range between 0 and 5, and the best-fitting parameters $n$
and $R_{e,n}$ are used both to describe the scale-lengths of the
galaxies and to interpret their dynamical properties, as discussed in
Section 6.

Derived values of $n$ and $R_{e,n}$ for galaxies in our sample are
reported in Table 1. We note that even allowing the Sersic index
to range out to n=8 the derived n are then found well below n=5.8. \\
We model each galaxy central PSF component with analytic functions
derived from visually selected stars in the surrounding ($30'' \times
30''$) region of each galaxy.  A 2D radial multi-gaussian function has
been fitted to tens of selected stars around each galaxy and outputs
have been stacked together to provide an appropriate PSF image (for
each galaxy) to be convolved to the galaxy best-fit 2D model to better
reproduce each observed galaxy light profile. In this way, we also
account for PSF variations over the ACS field.  More details of our
approach to modelling \textit{HST}/ACS galaxy morphologies in the
$1.0<z<1.5$ range will be given elsewhere (Rettura et al., in
preparation). The reliability and the accuracy of the
morphological analysis from GOODS ACS data in the range of magnitudes and sizes
probed by our sample is extensively discussed in \citet{Ravin06}. \\ In
addition, as the sample we use was mainly spectroscopically and color
selected, a visual analysis has also been performed to define a visual
morphological classification. We identified 37 ellipticals (E), 5
lenticulars (S0) and 6 bulge-dominated spirals (Sa/Sab; see bottom
panel of Fig.  \ref{zspec}).  Images of the 6 early-type-spiral
galaxies are shown in Fig.~\ref{spirals}. We also note that
CL1252-4419 and CL1252-4420 are the central Bright Cluster Galaxies
(BCGs) of CL1252 and are thought to be in mutual dynamical interaction
\citep{Blake03}. As shown in Fig. ~\ref{core}, the subtraction of our
S\'ersic 2D-models for these two galaxies reveals evidence of such an
interaction, in the form of an S-shaped residual that links the two
galaxy centers. As shown in Fig. ~\ref{ncolor}, we also find that
$n<3$ systems tend to have bluer $U-I$ rest-frame colors resulting in
a bimodal distribution which is similar to what has been found shown
by several authors (e.g. \citet{Kauffmann03}).
   \begin{figure}
   \centering
   \includegraphics[width=8cm]{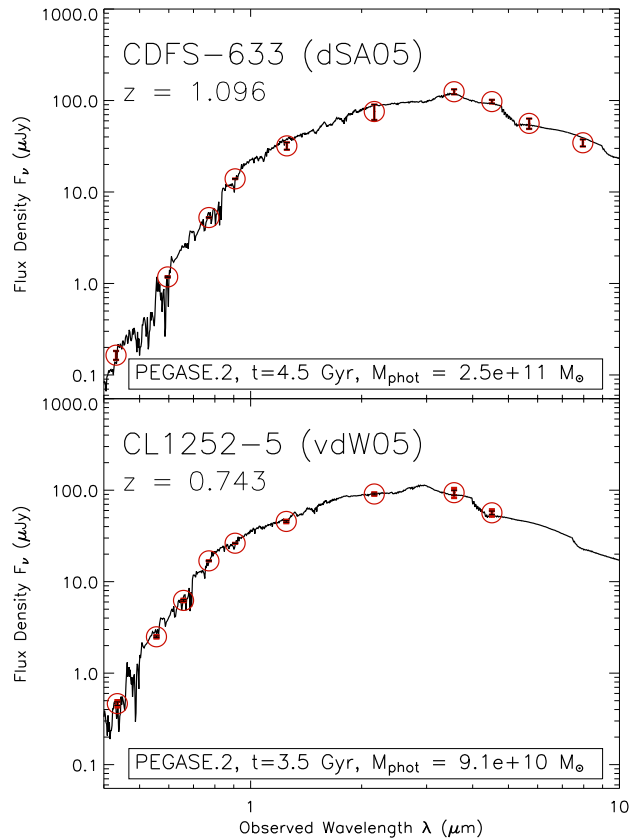}
      \caption{SEDs of an early type galaxies at $z \sim 1$ in each of
               our fields.  The red open circles, with
               corresponding error bars, are observed flux densities
               in the \textit{HST}/ACS ($B_{F435W}$, $V_{F606W}$,
               $i_{775W}$, $z_{F850lp}$),\textit{VLT}/ISAAC
               ($J$,$K_s$) and \textit{Spitzer}/IRAC ($3.6
               \mu\mbox{}m$, $4.5 \mu\mbox{}m$, $5.8 \mu\mbox{} m$,
               $8.0 \mu\mbox{}m$) passbands (top panel) and
               \textit{VLT}/FORS2 ($B$, $V$, $R$), \textit{HST}/ACS
               ($i_{775W}$, $z_{F850lp}$), \textit{VLT}/ISAAC
               ($J_s$,$K_s$) and \textit{Spitzer}/IRAC ($3.6
               \mu\mbox{}m$, $4.5 \mu\mbox{}m$) passbands (bottom
               panel). As an illustration, best-fit PEGASE.2 models
               are also shown (black solid line). The best-fit mass
               and age estimates are reported in each panel. Average
               errors on ages are $\pm 1$ Gyr, errors on masses are
               $\sim 40\%$ (i.e., $0.15$ dex).  }
      \label{sedfit}
   \end{figure}

   \begin{figure}[h!]
   \centering
   \includegraphics[width=8cm]{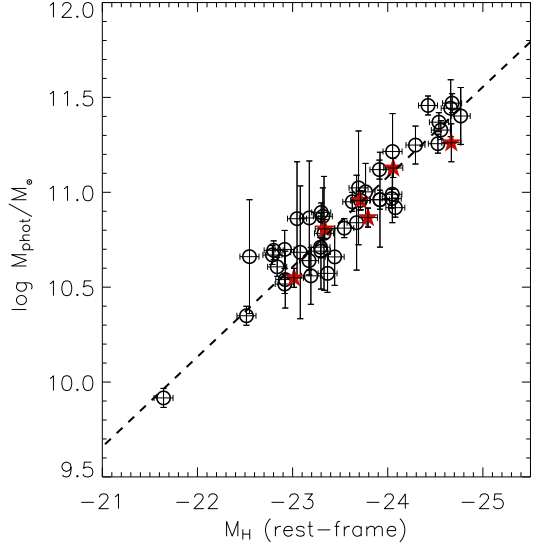}
     \caption{Each galaxy best-fit model rest-frame $H$-band ($\sim
     1.6 \mu m$) absolute magnitude, $M_H$, is plotted against derived
     photometric-stellar mass.  The dashed line correspond to a
     straight-line fit to the data (see text).  Filled red stars are
     used to indicate the bulge-dominated spirals sub-sample.}
         \label{Hmass}
   \end{figure}

\section{Photometric-stellar masses at $z\sim 1$}

\subsection{Stellar mass determination}
We derive stellar masses for each galaxy in our sample using
multiwavelength matched aperture photometry from 10 and 9 passbands
for the CDFS and CL1252 fields respectively. For each galaxy, we
compare the observed SED with a set of composite stellar populations
(hereafter, CSP) templates computed with PEGASE.2 models
\citep{Fioc97}. \\ In Figure ~\ref{sedfit} we show the SEDs of
two $z \sim 1$ early type galaxies of our sample. As an illustration
we over-plot the best-fit PEGASE.2 models from which the
photometric-stellar mass is estimated. \\ The adopted grid of star formation history (SFH)
scenarios has been shown to consistently reproduce observations of
galaxy counts at $z \sim 0$ froum UV to optical wavelengths
\citep{Fioc99}, and to result in reliable photometric redshifts with
the code Z-PEG \citep{LeBorgne02}. We assume a \citet{Kroupa01} IMF
and dust-free model templates. In the template SEDs used here, the
metallicity evolves consistently with the SFH: stars are gradually
formed with the same metallicity as the ISM, reaching about solar
values at the ages of interest in this study. We note that the
stellar masses computed in this paper corresponds to the mass
locked into stars plus the mass of the remnants (white dwarfs, 
neutron stars, black holes).

By comparing each observed SED with these atlases of synthetic
spectra, we construct a 3D $\chi^{2}$ space spanning a wide range of
SFHs, ages and stellar masses. We adopted same template
parameters as listed in Table 1 of \citet{LeBorgne02}. The age, SFH
scenario and stellar mass of the model giving the lowest $\chi^2$ are
taken as the best-fit estimates of the galaxy luminosity-weighted age,
SFH, and galaxy mass in stars.  These photometric-stellar mass
estimates take into account the evolution with galaxy age of the mass
fractions for each template as described in \citet{Rocca04} (see also
their Fig.  3).  The range of acceptable ages for a given galaxy has
been limited by the age of the universe at its observed redshift.\\
The errors on the ages and the masses are estimated by sampling the
full probability distribution (i.e. a function of the 3-dimension
space of free parameters). A 3D confidence region, around the measured
best-fit values, is set to contain 68.3 \% of the joint probability
distribution of the free parameters.  The errors correspond to the
projections of the confidence region for 3 interesting
parameters onto each free parameter axis. This procedure results in
typical errors of galaxy ages about 1 Gyr and typical uncertainties on
the mass determination of about $\sim 40\%$ (i.e., $0.15$ dex).  The
derived estimates of photometric-stellar masses for each galaxy are
summarized in Table 1. \\
It remains well known that the derived 
stellar masses depend on the adopted star formation history. \\ We have 
also investigated the effect of
dust extinction on the best-fit photometric-stellar masses by
including a fourth free parameter, $0.0< E(B-V)< 0.4$, following the
\citet{Cardelli89} prescription.  By performing the fit on 28 galaxies
for which IRAC photometry is available in all 4 bands, we find that in
$ \sim 40\%$ of the cases $E(B-V)=0$ gives the best fit.  In the remaining
cases, masses which are lower by $0.2 \pm 0.1$ dex are found, with
corresponding $E(B-V) \le 0.2$. This test supports the validity of the
dust-free model assumption, as also widely used in the literature for
early-type galaxies.
\\ Many authors have pointed
out that the use of the near-IR luminosity may allow a more secure
determination of stellar masses, as IRAC photometry samples the
rest-frame near IR-domain in this particular redshift range.  This
corresponds to $H$-band rest-frame wavelengths where the galaxy
luminosity is dominated by the old stars and so is expected to be
strongly correlated with the underlying photometric-stellar mass of
early-type galaxies.  To illustrate this point, we computed the
rest-frame $H$-band ($\sim 1.6 \mu m$) absolute magnitude, $M_H$, for
our $z \sim 1$ early-type galaxy sample, by applying an $H$-band
filter transmission function to each best-fit model.  In Figure
~\ref{Hmass}, we plot $M_H$ against the derived stellar mass resulting
in a tight correlation.  The dashed line corresponds to a
straight-line fit to the data that yields:
\begin{equation}  \label{eq:hmass}
log\frac{M_{phot}}{M_{\odot}} = -((0.47\pm0.02) \cdot M_H + (0.3\pm0.1))
\end{equation}

\begin{figure}[h!]
   \centering
   \includegraphics[width=8cm]{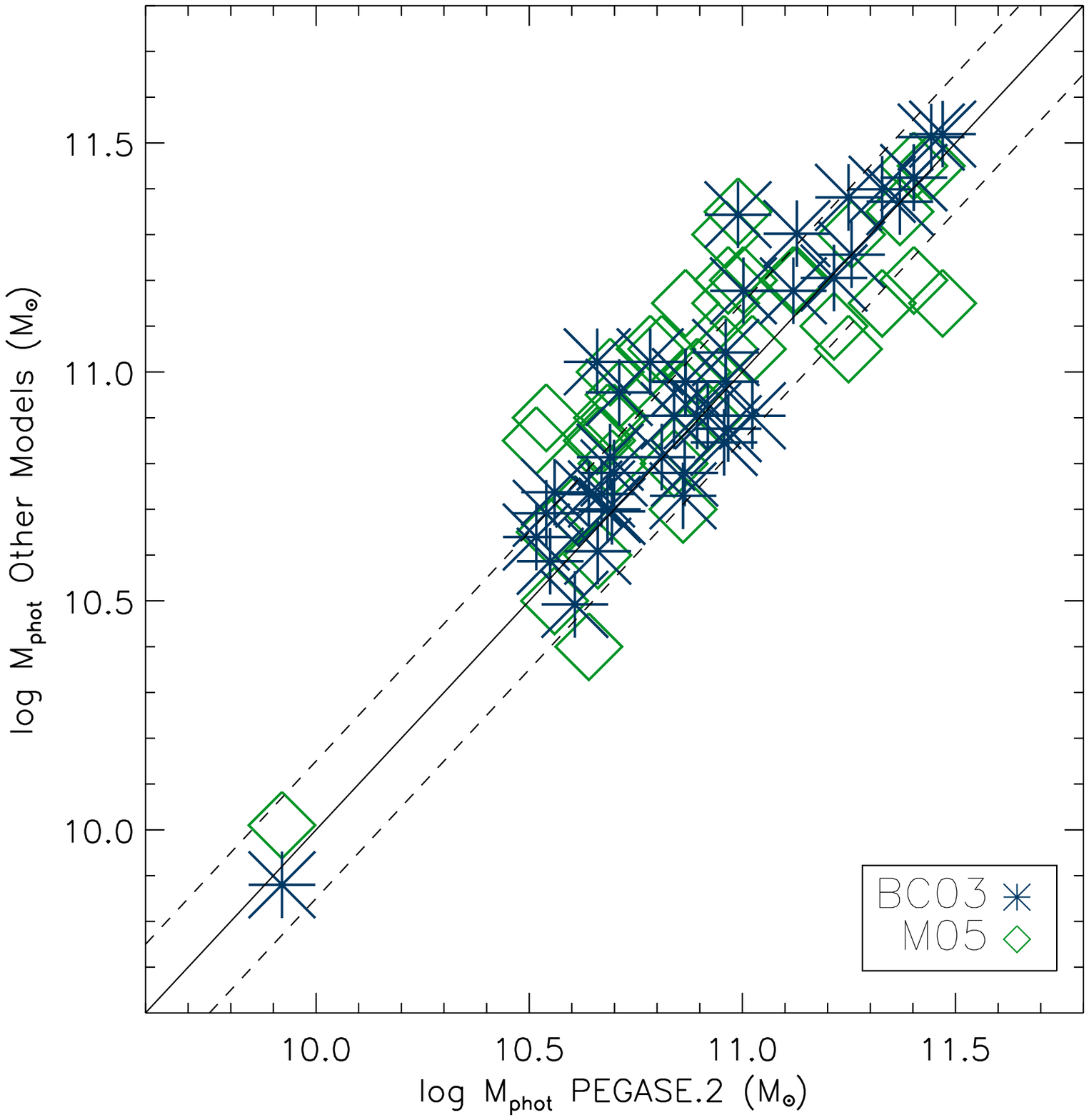}
   \includegraphics[width=8cm]{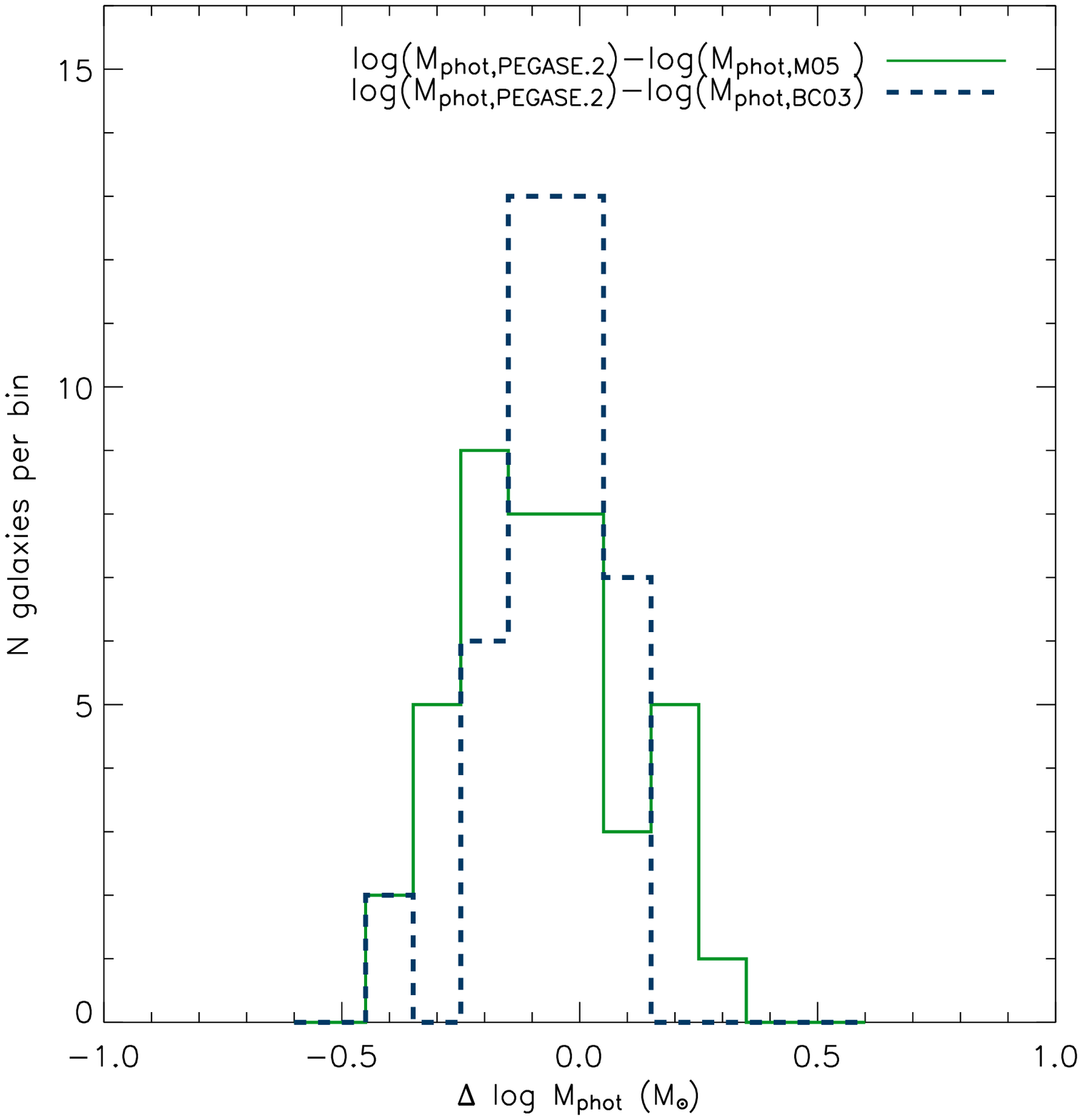}
      \caption{Top panel: Comparison of different model predictions in
      determining photometric-stellar masses of early-type galaxies at
      $z \sim 1$.  PEGASE.2 \citep{Fioc97} results are compared with
      \citet{Maraston05} and \citet{BC03} models, both with various
      exponentially-declining SFHs ($0.1 < \tau < 10 Gyr$), solar
      metallicity and \citet{Kroupa01} IMF (see text). The dashed lines
      indicate errors of $40 \%$.  Bottom panel: difference between
      PEGASE.2 and BC03 or M05 models mass estimates, $<\Delta
      log(M_{phot}/M_{\odot})>= -0.06,\, -0.08$ dex, and
      $\sigma_{\Delta log(M_{phot}/M_{\odot})}= 0.11,\, 0.17$ dex,
      respectively}
         \label{3models}
   \end{figure}

\subsection{Comparison with other stellar population synthesis models}

As stated above, our photometric-stellar masses are derived by
sampling the entire relevant wavelength domain of stellar light
emission, from the rest-frame UV through the NIR. However, the
reliability of spectrum synthesis models at $\lambda_{obs} \sim 2\mu
m$ has long been debated ( \citet{Maraston98} and references therein).

In the rest-frame near-IR regime, in early stages of the galaxy
evolution, a short-duration thermally pulsating (TP-) AGB phase is
known to be relevant. The PEGASE.2 models we primarily use in our work
compute isochrones up to the TP-AGB phase using the equations proposed
by \citet{Groen93} (see also \citet{Fioc96}). \\ Measuring
the evolution of the rest-frame K-band FP from $z \sim 1$ to the present,
\citet{vdw05b} find the evolution in $M/L_K$ to be slower than the
evolution in $M/L_B$, as expected from stellar populations models.
This study also finds the dust-free \citet{Maraston05} (M05) models
(which implement the short-duration TP-AGB phases adopting a `fuel
consumption' approach) to provide a better fit to the data than
dust-free \citet{BC03} (BC03) models (which implement the TP-AGB
phases with an empirical prescription).\\ \citet{vdw05b} also suggest
that the models uncertainties caused by the different treatment of AGB
stars can severely hamper the determination of $z \sim 1$ early-type
galaxy masses from rest-frame near-IR photometry.\\

We compare our photometric-stellar mass estimates based on CSP
dust-free PEGASE.2 models with those obtained with a set of dust-free
CSPs models built with both \citet{Maraston05} and \citet{BC03}
models, using exponentially-declining SFHs ($0.1 < \tau < 10$ Gyr),
solar metallicity and \citet{Kroupa01} IMF. Since M05 models provide
calibrated spectra only up to 2.5 $\mu$m, when using M05 models, we
limit the SED fitting range to the 4.5 $\mu$m IRAC channel.  For our
sample, we find consistent photometric-stellar mass estimates with
CSPs PEGASE.2, M05 and BC03 models within typical errors of $40 \%$
(see top panel of Figure~\ref{3models}). The average difference
between the photometric-stellar masses estimated with PEGASE.2 and
BC03 or M05 model is $<\Delta log(M_{phot}/M_{\odot})> = -0.06$ dex
and $-0.08$ dex respectively, with a standard deviation of
$\sigma_{\Delta log(M_{phot}/M_{\odot})} = 0.11$ dex and $0.17$ dex
(see bottom panel of Figure~\ref{3models}).\\ Thus, recognizing the
uncertainties involved in the SED fitting technique and in the
multiwavelength photometry, we find the overall difference in
photometric-stellar mass estimates for early-type galaxies at $z \sim
1$ from PEGASE.2, BC03, M05 not to be significant.  This is not
surprising. In fact, regardless of the actual implementation of the
TP-AGB phase in the different codes, the rest-frame K-band is expected
to be dominated by AGB stars only for $0.1< T_{\rm AGB}/\mbox{Gyr} <
2.0$, while we find our sample ages all to be much greater than 2Gyr,
in agreement with similar studies of $z \sim 1.0$ early-type galaxies
( \citet{Cimatti02}, \citet{Franx03},
\citet{Cimatti04},\citet{Glaze04}, \citet{McCarthy04},
\citet{Fontana04}, \citet{Saracco04}, \citet{Rocca04}). These
results contrast with those of \citet{vdw05b}, however a direct
comparison is not possible since they adopted Simple Stellar Population
models and different estimates of the photometric errors (weights) on the IRAC photometry.\\

\section{Dynamical masses of early-type galaxies}

The central velocity dispersion measures the random motions
of the stars averaged along the line of sight through the galaxy, and
is known to be a good tracer of the dynamical state of spheroidal, non
rotationally-supported, galactic systems in virial equilibrium.  The
dynamical mass of a spherical, non-rotating, isotropic model with a
S\'ersic $R^{1/n}$ SB profile is given by \citet{Bertin02}:
\begin{equation}  \label{eq:massdynKV}
M_{dyn} = K_V(n,R_a) \frac{\sigma_a^2 R_{e,n}}{G},
\end{equation}
where $\sigma_a$ is the velocity dispersion measured within an
aperture of radius $R_a$, and $K_V(n,R_a)$ is the so-called `virial
coefficient', that takes into account the differences between the
virial and the effective radii, and between the virial and the
observed velocity dispersions.  If the galaxy SB profile is well
described by the de Vaucouleurs law ($n=4$ in the previous equation)
and if $\sigma_a$ is the central velocity dispersion (measured within
a tenth of the effective radius), the `classical' value $K_V\simeq 5$ (e.g.,
\citet{Michard80}) is recovered.  Using high signal-to-noise (S/N)
integral field spectroscopy of a sample of early-type galaxies at
$z\sim 0$ (the SAURON project; \citet{Bacon01} ), \citet{Cappellari05}
have recently found that the above expression with $n=4$ and $K_V=5.0\pm
0.1$ reproduces the galaxy dynamical masses, closely matching the
values obtained with a much more accurate modelling (the so-called
`Schwarzschild method').

We use Eq. (~\ref{eq:massdynKV}) with $K_V$ properly computed taking
into account our best-fitting values of $n$ and $R_{e,n}$ from the
\textit{HST}/ACS images, and the published, aperture-corrected
velocity dispersions (following the prescriptions of \citet{Bertin02}
for apertures of radius $R_J/R_{e,n}$).  Note that the mass computed
by such a relation corresponds to the galaxy \emph{total} mass, and it
is appropriate both for a single component (stars only), and for a
two-component (stars and dark matter) system, provided that the dark
matter density distribution parallels the stellar one (see
\citet{Lanzoni03}). The values of each galaxy $K_V$ and dynamical
mass, $M_{dyn}(K_V)$ are presented in Table 1.
\begin{table*}
\caption{The early-type galaxy sample at $z \sim 1$. IDs are from the
original surveys and the reader is referred to the aforementioned
papers for objects selection and coordinates. $z_{\textit{F850lp}}$ is
the $3\arcsec$ aperture magnitude in the \textit{HST}/ACS $F850lp$
filter; $M_H$ is the rest-frame absolute magnitude obtained applying
an $H$-band standard filter transmission function to each best-fit
model; $U-V$ is a rest-frame color. The magnitudes and the color
are in the AB system. Only statistical errors are quoted for photometric-stellar masses.} 
\label{table:1} \centering
\begin{tabular}{l l l c l c c c c c c}     
\hline\hline
\\
ID & $zspec$ & $z_{\textit{F850lp}}$ & $M_H$ & $U-V$ & $n$ & $R_{e,n}$ & $K_V$ & $\sigma_0$ & $log\frac{M_{dyn}(K_V)}{M_{\odot}}$ & $log\frac{M_{phot}}{M_{\odot}}$ \\
   &  & (mag) & (mag) & & & (kpc) & & (km/s) & &\\
\hline
      CDFS-1   &      1.089   &      21.18   &     -24.05   &    1.68      &       3.63$\pm$0.25   &       3.17$\pm$0.12   &       5.15$\pm$0.21   &   231$\pm$	15   &      11.31$\pm$0.08   &      10.99$\pm$0.15\\
      CDFS-2   &      0.964   &      20.75   &     -23.91   &    1.80      &       5.00$\pm$0.32   &       5.45$\pm$0.21   &       3.90$\pm$0.22   &   200$\pm$9   &      11.30$\pm$0.07   &      11.12$\pm$0.05\\
      CDFS-3   &      1.044   &      21.78   &     -23.33   &    1.73      &       4.11$\pm$0.37   &       1.07$\pm$0.05   &       5.61$\pm$0.24   &   300$\pm$	30   &      11.10$\pm$0.11   &      10.78$\pm$0.30\\
      CDFS-4   &      0.964   &      20.66   &     -24.66   &    1.99      &       4.22$\pm$0.22   &       8.42$\pm$0.53   &       4.47$\pm$0.22   &   336$\pm$	18   &      11.99$\pm$0.09   &      11.44$\pm$0.15\\
      CDFS-5   &      0.685   &      21.31   &     -22.80   &    1.99      &       2.63$\pm$0.11   &       1.79$\pm$0.09   &       6.24$\pm$0.09   &   194$\pm$	15   &      10.99$\pm$0.09   &      10.69$\pm$0.05\\
      CDFS-6   &      0.660   &      20.58   &     -22.79   &    1.81      &       5.00$\pm$0.39   &       1.81$\pm$0.18   &       4.55$\pm$0.23   &   208$\pm$9   &      10.92$\pm$0.09   &      10.67$\pm$0.05\\
      CDFS-7   &      1.135   &      21.43   &     -24.53   &    1.93      &       5.00$\pm$0.33   &       3.18$\pm$0.34   &       3.76$\pm$0.21   &   232$\pm$	19   &      11.18$\pm$0.12   &      11.26$\pm$0.05\\
      CDFS-8   &      1.125   &      22.21   &     -23.18   &    1.87      &       4.99$\pm$0.17   &       2.07$\pm$0.18   &       4.45$\pm$0.10   &   253$\pm$	70   &      11.14$\pm$0.22   &      10.86$\pm$0.30\\
      CDFS-9   &      1.097   &      21.78   &     -23.92   &    1.65      &       5.00$\pm$0.25   &       2.08$\pm$0.10   &       4.44$\pm$0.14   &   215$\pm$	45   &      11.00$\pm$0.18   &      10.96$\pm$0.25\\
     CDFS-10   &      1.119   &      22.08   &     -23.30   &    1.68      &       4.44$\pm$0.29   &       0.62$\pm$0.05   &       6.06$\pm$0.19   &   275$\pm$	49   &      10.82$\pm$0.17   &      10.69$\pm$0.20\\
     CDFS-11   &      1.096   &      22.07   &     -23.77   &    1.96      &       5.00$\pm$0.14   &       2.39$\pm$0.18   &       4.07$\pm$0.09   &   208$\pm$	33   &      10.99$\pm$0.15   &      11.00$\pm$0.15\\
     CDFS-12   &      1.123   &      21.80   &     -23.69   &    1.85      &       5.00$\pm$0.31   &       1.44$\pm$0.11   &       4.76$\pm$0.18   &   262$\pm$	20   &      11.04$\pm$0.10   &      11.02$\pm$0.30\\
     CDFS-13   &      0.980   &      20.76   &     -24.05   &    1.87      &       5.00$\pm$0.15   &       2.58$\pm$0.24   &       4.02$\pm$0.10   &   247$\pm$	10   &      11.17$\pm$0.08   &      11.21$\pm$0.20\\
     CDFS-14   &      0.984   &      21.59   &     -23.71   &    1.65      &       5.00$\pm$0.18   &       4.80$\pm$0.54   &       3.95$\pm$0.12   &   197$\pm$	21   &      11.23$\pm$0.13   &      10.96$\pm$0.05\\
     CDFS-15   &      0.622   &      20.77   &     -22.92   &    1.98      &       5.00$\pm$0.21   &       3.53$\pm$0.22   &       4.09$\pm$0.13   &   317$\pm$	21   &      11.53$\pm$0.09   &      10.70$\pm$0.10\\
     CDFS-16   &      0.669   &      20.70   &     -23.30   &    1.99      &       5.00$\pm$0.11   &       2.85$\pm$0.31   &       4.03$\pm$0.06   &   262$\pm$	36   &      11.26$\pm$0.15   &      10.89$\pm$0.05\\
     CDFS-17   &      0.954   &      21.39   &     -23.71   &    1.64      &       2.91$\pm$0.31   &       4.65$\pm$0.13   &       5.84$\pm$0.36   &   305$\pm$	31   &      11.77$\pm$0.11   &      10.96$\pm$0.05\\
     CDFS-20   &      1.022   &      21.26   &     -24.29   &    1.98      &       4.84$\pm$0.28   &       3.50$\pm$0.08   &       4.20$\pm$0.19   &   199$\pm$	15   &      11.13$\pm$0.09   &      11.25$\pm$0.10\\
     CDFS-21   &      0.735   &      21.07   &     -22.51   &    1.63      &       5.00$\pm$0.31   &       1.05$\pm$0.10   &       4.90$\pm$0.18   &   149$\pm$8   &      10.42$\pm$0.09   &      10.35$\pm$0.05\\
     CDFS-22   &      0.735   &      20.68   &     -23.63   &    1.71      &       5.00$\pm$0.25   &       5.88$\pm$0.31   &       3.79$\pm$0.17   &   225$\pm$	11   &      11.42$\pm$0.08   &      10.95$\pm$0.05\\
     CDFS-23   &      1.041   &      22.04   &     -23.05   &    1.90      &       4.01$\pm$0.27   &       2.74$\pm$0.22   &       4.91$\pm$0.20   &    70$\pm$	15   &      10.18$\pm$0.19   &      10.86$\pm$0.30\\
     CDFS-24   &      1.042   &      21.10   &     -24.06   &    1.79      &       2.25$\pm$0.11   &       9.13$\pm$0.35   &       6.79$\pm$0.18   &   210$\pm$	16   &      11.80$\pm$0.09   &      11.13$\pm$0.05\\
     CDFS-25   &      0.967   &      21.55   &     -23.32   &    1.85      &       5.00$\pm$0.39   &       1.20$\pm$0.23   &       4.95$\pm$0.23   &   258$\pm$	18   &      10.96$\pm$0.14   &      10.87$\pm$0.15\\
     CDFS-26   &      1.129   &      21.24   &     -24.67   &    1.65      &       4.18$\pm$0.38   &      11.17$\pm$0.64   &       4.55$\pm$0.42   &   249$\pm$	25   &      11.86$\pm$0.13   &      11.26$\pm$0.10\\
     CDFS-27   &      1.128   &      21.95   &     -23.37   &    1.42      &       5.00$\pm$0.27   &       7.84$\pm$0.45   &       3.76$\pm$0.20   &   135$\pm$	30   &      11.10$\pm$0.19   &      10.57$\pm$0.10\\
     CDFS-28   &      0.954   &      21.85   &     -23.34   &    1.65      &       1.53$\pm$0.26   &       2.29$\pm$0.21   &       7.08$\pm$0.15   &   445$\pm$	84   &      11.87$\pm$0.17   &      10.81$\pm$0.05\\
     CDFS-29   &      1.128   &      21.07   &     -24.42   &    1.92      &       5.00$\pm$0.31   &       2.39$\pm$0.13   &       4.34$\pm$0.19   &   221$\pm$	17   &      11.07$\pm$0.10   &      11.46$\pm$0.05\\
    CL1252-1   &      0.671   &      20.61   &     -23.29   &    1.65      &       4.82$\pm$0.18   &       4.27$\pm$0.54   &       4.04$\pm$0.12   &   219$\pm$	12   &      11.28$\pm$0.10   &      10.71$\pm$0.05\\
    CL1252-2   &      0.658   &      20.88   &     -22.55   &    1.90      &       4.23$\pm$0.11   &       1.23$\pm$0.07   &       5.39$\pm$0.07   &   216$\pm$6   &      10.86$\pm$0.05   &      10.66$\pm$0.30\\
    CL1252-3   &      0.844   &      20.46   &     -24.08   &    1.49      &       1.59$\pm$0.15   &       4.75$\pm$0.37   &       7.14$\pm$0.12   &   166$\pm$7   &      11.34$\pm$0.07   &      10.92$\pm$0.05\\
    CL1252-4   &      0.743   &      20.61   &     -23.01   &    1.63      &       2.50$\pm$0.33   &       4.37$\pm$0.27   &       6.31$\pm$0.40   &   202$\pm$8   &      11.42$\pm$0.08   &      10.55$\pm$0.05\\
    CL1252-5   &      0.743   &      20.35   &     -23.92   &    1.65      &       5.00$\pm$0.23   &       4.91$\pm$0.31   &       3.94$\pm$0.15   &   251$\pm$9   &      11.45$\pm$0.07   &      10.96$\pm$0.05\\
    CL1252-6   &      0.734   &      20.68   &     -23.54   &    1.65      &       4.89$\pm$0.26   &       1.92$\pm$0.16   &       4.57$\pm$0.16   &   211$\pm$5   &      10.96$\pm$0.07   &      10.81$\pm$0.05\\
    CL1252-7   &      0.753   &      20.16   &     -24.04   &    1.57      &       3.77$\pm$0.31   &       2.43$\pm$0.19   &       5.16$\pm$0.23   &   213$\pm$5   &      11.12$\pm$0.07   &      10.97$\pm$0.05\\
    CL1252-8   &      1.069   &      21.98   &     -23.17   &    1.68      &       5.00$\pm$0.19   &       2.71$\pm$0.25   &       4.25$\pm$0.12   &    63$\pm$	13   &      10.03$\pm$0.19   &      10.64$\pm$0.05\\
    CL1252-9   &      1.036   &      22.20   &     -22.92   &    1.57      &       4.24$\pm$0.29   &       1.42$\pm$0.06   &       5.24$\pm$0.19   &   102$\pm$	16   &      10.25$\pm$0.14   &      10.52$\pm$0.05\\
    CDFS-369   &      0.894   &      22.31   &     -21.64   &    1.35      &       1.85$\pm$0.39   &       0.84$\pm$0.05   &       7.36$\pm$0.20   &   119$\pm$	21   &      10.31$\pm$0.16   &       9.92$\pm$0.05\\
    CDFS-467   &      0.895   &      21.54   &     -22.92   &    1.68      &       3.45$\pm$0.39   &       1.52$\pm$0.10   &       5.72$\pm$0.28   &   140$\pm$	18   &      10.60$\pm$0.14   &      10.54$\pm$0.15\\
    CDFS-532   &      1.215   &      22.13   &     -23.67   &    1.68      &       2.04$\pm$0.33   &       1.41$\pm$0.10   &       6.82$\pm$0.23   &   260$\pm$	30   &      11.18$\pm$0.13   &      10.84$\pm$0.25\\
    CDFS-547   &      1.222   &      22.03   &     -23.19   &    1.50      &       1.90$\pm$0.31   &       0.74$\pm$0.04   &       7.50$\pm$0.16   &   256$\pm$	28   &      10.93$\pm$0.11   &      10.56$\pm$0.15\\
    CDFS-571   &      0.955   &      21.22   &     -23.79   &    1.57      &       5.00$\pm$0.15   &       6.81$\pm$0.48   &       3.83$\pm$0.11   &   182$\pm$	21   &      11.30$\pm$0.12   &      10.87$\pm$0.05\\
    CDFS-590   &      1.222   &      22.71   &     -23.08   &    1.73      &       4.99$\pm$0.25   &       4.09$\pm$0.52   &       4.03$\pm$0.16   &   119$\pm$	49   &      10.73$\pm$0.30   &      10.68$\pm$0.35\\
    CDFS-633   &      1.096   &      21.04   &     -24.77   &    1.96      &       4.78$\pm$0.21   &       6.66$\pm$0.12   &       4.00$\pm$0.16   &   260$\pm$	23   &      11.62$\pm$0.09   &      11.40$\pm$0.15\\
    CDFS-354   &      0.667   &      21.09   &     -22.84   &    1.65      &       3.58$\pm$0.28   &       3.08$\pm$0.12   &       5.20$\pm$0.24   &    99$\pm$	19   &      10.56$\pm$0.17   &      10.61$\pm$0.05\\
 CL1252-6106   &      1.235   &      21.40   &     -24.56   &    1.79      &       3.38$\pm$0.20   &       2.76$\pm$0.35   &       5.42$\pm$0.17   &   294$\pm$	10   &      11.48$\pm$0.09   &      11.33$\pm$0.05\\
 CL1252-9077   &      1.241   &      22.06   &     -23.44   &    1.50      &       2.36$\pm$0.21   &       3.15$\pm$0.11   &       6.38$\pm$0.23   &   130$\pm$	14   &      10.90$\pm$0.11   &      10.66$\pm$0.15\\
 CL1252-4419   &      1.238   &      21.32   &     -24.54   &    1.81      &       1.86$\pm$0.10   &       8.56$\pm$0.84   &       7.72$\pm$0.12   &   302$\pm$	24   &      12.15$\pm$0.10   &      11.37$\pm$0.05\\
 CL1252-4420   &      1.234   &      21.37   &     -24.68   &    1.86      &       4.68$\pm$0.36   &       9.82$\pm$0.77   &       4.07$\pm$0.32   &   323$\pm$	21   &      11.99$\pm$0.11   &      11.47$\pm$0.05\\

\hline
\end{tabular}
\end{table*}


   \begin{figure}[h!]
   \centering
   \includegraphics[width=8cm]{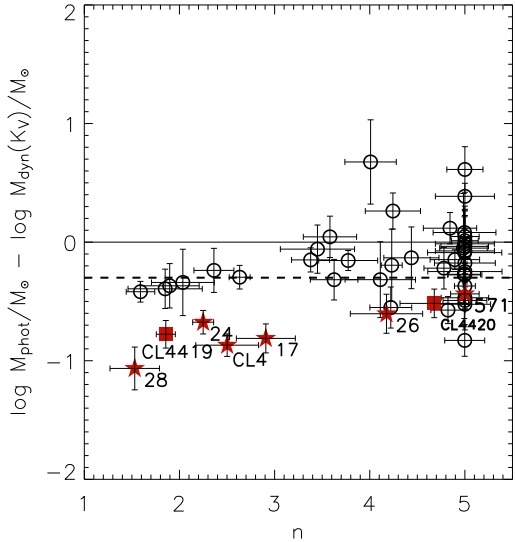}
      \caption{Photometric-stellar vs dynamical mass differences,
         $log(M_{phot})-log(M_{dyn}(K_V))$, as a function of the
         S\'ersic index, $n$. \citet{Kroupa01} IMF is
         assumed. Adopting a \citet{Salpeter55} IMF all the $log
         M_{phot}$ values would constantly increase by $\approx 0.3$
         dex. This effect could be visualized by shifting the
         one-to-one line (the horizontal solid line) of $0.3$ dex
         downward, to the position of the dashed line. Symbols are the
         same as in Fig.  ~\ref{ncolor} }
         \label{deltasersic}
   \end{figure}



    \begin{figure}[t!]
    \centering
    \includegraphics[width=8cm]{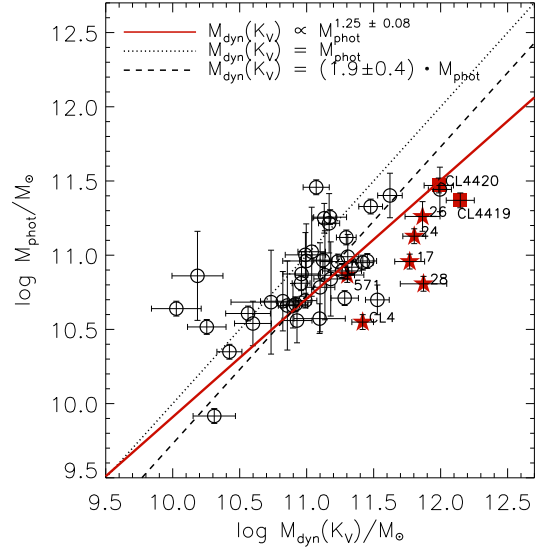}
      \caption{Total dynamical, $M_{dyn}(K_V)$,
       vs. photometric-stellar, $M_{phot}$, mass estimates for the
       entire early-type galaxy sample at $z \sim 1$. Symbols are the
       same as in Figure ~\ref{ncolor}. The dashed lines
       indicates the $1.5$-sigma clipped mean offset of $-0.27 \pm
       0.09$ dex.  The red dotted-dashed line show the best fit
       correlation between dynamical mass and photometric-stellar
       masses (see text).  }
    \label{compmass1}
   \end{figure}

\section{Comparing photometric-stellar and dynamical mass estimates}

Before discussing the direct comparison between photometric-stellar masses and total dynamical masses using Eq.(\ref{eq:massdynKV}), we describe
some possible biases that may in principle affect both measurements.\\
As mentioned above, using the dynamical mass estimates rely on the simple assumption
that an early-type galaxy is a spherical, isotropic, non
rotationally-supported galactic system in virial equilibrium. The
presence of a rotational component in the form of a disk or the
presence of dynamical galaxy-to-galaxy interaction may have an effect
which is not easy to quantify.\\

In Figure ~\ref{deltasersic}, we plot mass differences for our sample,
$log(M_{phot})-log(M_{dyn}(KV))$, against the S\'ersic index, $n$.  We
generally find the mass deviation to be larger for $n<3$ galaxies. \\
The modulus of the differences of the bulgy-spiral sub-sample of Fig.
~\ref{spirals} (indicated as filled red stars in Fig. ~\ref{deltasersic})
is larger than the average at any $n$; with dynamical masses being
larger than photometric-stellar masses with decreasing S\'ersic
indices. For these galaxies, the flattening of the SB profile reveals
an increasing contribution of the disk component. As a result, their
measured velocity dispersions, and thus the derived dynamical mass
estimates, can be biased to larger values because of the presence of a
rotating stellar disk which is generally included in the spectroscopic
apertures ($ 1\arcsec$) at these redshifts. This trend is also seen in 
recent observations by \citet{Ganda05} of local
spiral-galaxies using SAURON integral field spectroscopy. Their
measured stellar kinematic maps often show a central depression in the
velocity dispersion, with velocity dispersion profiles increasing
outwards. However, one would also expect this effect to be maximum in
case of edge-on galaxies and to be minimum/negligible for face-on
galaxies. In our case, this would imply that at least for the face-on
objects CDFS-24, CDFS-26, CDFS-571 and CL1252-4, the measured velocity
dispersion is probably slightly affected by the presence of a disk.\\
Among the galaxies with dynamical masses significantly exceeding the
photometric ones, we note in Fig. ~\ref{deltasersic} the two
elliptical galaxies, CL$4419$ and CL$4420$ (filled red squares).
These are the central Bright Cluster Galaxies (BCGs) of CL1252 which
show signs of mutual dynamical interaction (see Fig. ~\ref{core}).
Such interactions may, in principle, introduce
anisotropies into the motion of their stars, thus affecting the
$\sigma_a$ measurements in a way that is not trivial.  We also note
that using field early-type galaxies and cluster BCGs in our study is
justified by the results of \citet{vdw05a}, who find that the most massive
galaxies ($M \gtrsim 10^{11} M_{\odot}$) lie on the same Fundamental
Plane line regardless of their environments.\\

Photometric stellar mass estimates may also be affected by
systematics.  First, photometric-stellar masses weakly depend on the
assumed model of dust extinction and metallicity evolution.  Second
and more importantly, they depend on the assumed IMF.  For instance,
by using a \citet{Salpeter55} IMF, instead of our adopted
\citet{Kroupa01} IMF, all the $log M_{phot}$ values would be shifted
upwards by $\approx 0.3$ dex. In Fig. ~\ref{deltasersic}, this effect
is shown by shifting the one-to-one line (the horizontal solid line)
$0.3$ dex downward, to the position of the dashed line. Adopting a
\citet{Salpeter55} IMF a large number of galaxies would have
photometric stellar masses greater than the dynamical ones, in
agreement with the studies of
\citet{Cappellari05} and \citet{Ferreras05} in local early-type
galaxies, which also favour a \citet{Kroupa01} IMF. However, it is clear that 
our data cannot be used to constrain the choice of the IMF. \\

In Fig. ~\ref{compmass1}, we present the direct comparison of total
dynamical and photometric-stellar masses. We find that, adopting a
\citet{Kroupa01} IMF, the photometric-stellar masses reproduce the
dynamical ones with a $1.5$-sigma clipped mean offset of $-0.27 \pm
0.09$ dex (dashed line). This relation implies $M_{dyn}(K_V) =
(1.9\pm0.4) \cdot M_{phot}$ and it is illustrated in In
Fig. ~\ref{compmass1}. Therefore, if the IMF is not varying, this
first result suggests the presence of a $40-50 \%$ dark matter
component beyond several $R{e,n}$. \\ We also compute (dotted-dashed
red line) the best fit relation for the entire sample, which
yields:
\begin{equation}  \label{eq:massesbestfit}
M_{dyn}(K_V) = 10^{(-2.4 \pm 0.9)} \cdot M_{phot}^{(1.25 \pm 0.08)}.
\end{equation}
We note that the best-fit slope does not change if the
bulge-dominated spirals (red stars) are excluded from the fit. We have
also note that the relatively small error bar for the galaxy with $log
M_{phot} < 10$, does not have a significant impact on the
determination of the best-fit slope either.\\ Our finding is in
agreement with the picture of early-type galaxies, being dark matter
dominated ($M_{dyn} >> M_{phot}$) in the most massive systems and
baryon dominated ($M_{dyn} \gtrapprox M_{phot}$) in less massive
systems, on scale of several $R_e$. The power law index in Eq.
(\ref{eq:massesbestfit}) is indeed consistent with completely
independent estimates (at $ 0.3 < z < 1.0$) of total mass using strong
gravitational lensing by \citet{Ferreras05}. Accordingly, these
authors find the dark matter dominating in massive elliptical galaxies
while the stellar content dominates the mass budget in lower mass
galaxies. In this context, the evolution of the dark matter fraction
with the early-type galaxy mass has also been commonly invoked as a
possible explanation of the so-called `tilt' of the FP (e.g.,
\citet{Ferreras00}).

\section{Conclusions}

In this work, we have used photometry from 9-10 passbands to build
accurate SEDs, covering the 0.2-4$\mu$m rest-frame wavelengths, of a
sample of 48 early-type galaxies at $z\sim\! 1$ with published
velocity dispersions.  The large wavelength baseline and accuracy of
our photometric measurements allows us to compare measured stellar
masses with different spectrum synthesis models.\\ Based on our
sample, which spans a limited mass range $\log M_{phot} \simeq
[10,11.5]$, corresponding to the bright end of the stellar mass
function, $M \gtrsim M\ast$, we find that photometric-stellar mass
estimates are not strongly dependent on the choice of the stellar
population model. Regardless of the actual implementation of the
TP-AGB phase in the different codes, we find the overall difference in
photometric-stellar masses of early-type galaxies at $z \sim 1$ from
PEGASE.2, BC03, M05 not to be statistically significant (below 0.1 dex). 
\\  We have also investigated the other inherent systematic uncertainties 
on stellar masses, such as those due to reddening and adopted IMF, and found 
them of the order 0.2-0.3 dex.\\ We have
then compared our photometric-stellar masses to the total dynamical
masses as inferred from velocity dispersion measurements and
half-light radii measured in deep \textit{HST}/ACS images. Strong
deviations from $M_{dyn}=M_{phot}$ may be ascribed to possible biases
in dynamical mass measurements, as suggested by the evidence that
deviations increase for early-type galaxies with small disk components
and/or complex morphologies, as well as in galaxies showing signs of
dynamical interaction with close-by companions. In the other hand,
photometric-stellar masses depend on the assumed model of dust
extinction and metallicity evolution as well as from the assumed
IMF. \\ Assuming \citet{Kroupa01} IMF, we find the photometric-stellar
masses to reproduce the dynamical mass estimates with an an average
offset of 0.27 dex. We note that the average offset depends on the
assumed IMF, although a \citet{Salpeter55} IMF would produce
unphysical results by implying that a large number of galaxies would
have photometric-stellar masses greater than dynamical estimates.  \\
We also find that an increasing dark matter fraction with the
increasing total galaxy mass may be needed to explain the observed
trend in $M_{dyn}(K_V) \propto M_{phot}^{(1.25 \pm 0.08)}$.  It is
reassuring that a similar relation and slope was found in studies
where the galaxy mass was directly derived from strong lensing
models. The increase in $M_{dyn}/M_{phot}$ with increasing dynamical
mass that we find is consistent with the increase of the mass-to-light
ratio ($M_{dyn}/L$) of early type galaxies implied by FP studies. \\

We conclude that the determination of photometric stellar-masses of
massive early-type galaxies at $z\sim\! 1$ is robust against stellar
population models, when a large wavelength baseline is available, and
its accuracy hinges primarily on the adopted IMF. \\ Although a clear
correlation $M_{dyn}-M_{phot}$ is found, it remains difficult to use
photometric masses as reliable surrogates of total galaxy masses, or
mass-to-light ratios, over a wide mass range. Specifically, it
would be interesting to extend our study to less massive galaxies ($ M_{phot} <
10^{10} M\odot$) with possibly younger ages, not probed by our
sample. Further studies on galaxy-scale lensing systems (e.g. ,
\citep{Koopmans03}), with masses accurately determined by parameter
free strong lensing models, will surely stimulate significant progress
in the years to come.

\begin{acknowledgements}
      A.R. is very grateful to A. Renzini, G. Zamorani, J. Vernet and
      M. Pannella for useful discussions.\\ RAEF is affiliated to the
      Research and Science Support Department of the European Space
      Agency. The work of DS was carried out at Jet Propulsion
      Laboratory, California Institute of Technology, under a contract
      with NASA. The work by SAS was performed under the auspices of
      the U.S.  Department of Energy, National Nuclear Security
      Administration by the University of California, Lawrence
      Livermore National Laboratory under contract No. W-7405-Eng-48.
\end{acknowledgements}

\bibliography{5273}

\end{document}